\documentclass{ws-procs9x6-cpt19}
\begin{document}

\newcommand{\refeq}[1]{(\ref{#1})}
\def\etal {{\it et al.}}

\title{Cryogenic Penning-Trap Apparatus for Precision Experiments with Sympathetically Cooled (anti)protons}

\author{
	M.\ Niemann,$^1$ 
	T.\ Meiners,$^1$ 
	J.\ Mielke,$^1$ 
	N.\ Pulido,$^1$ 
	J.\ Schaper,$^{5,1}$ 
	M.J.\ Borchert,$^{5,1}$\\
    J.M.\ Cornejo,$^1$ 
    A.-G.\ Paschke,$^{1,2}$ 
    G.\ Zarantonello,$^{1,2}$ 
    H.\ Hahn,$^{2,1}$ 
    T.\ Lang,$^{1}$\\
    C.\ Manzoni,$^3$ 
    M.\ Marangoni,$^3$ 
    G.\ Cerullo,$^3$ 
    U.\ Morgner,$^1$ 
    J.-A.\ Fenske,$^2$\\
    A.\ Bautista-Salvador,$^{2,1}$ 
    R.\ Lehnert,$^{4,1}$ 
    S.\ Ulmer,$^5$ and 
    C.\ Ospelkaus$^{1,2}$\\
    }

\address{$^1$Institut f\"ur Quantenoptik, Leibniz Universit\"at Hannover,\\ 
	Welfengarten 1, 30167 Hannover, Germany, and\\
	Laboratorium f\"ur Nano- und Quantenengineering, Leibniz Universit\"at Hannover,\\ Schneiderberg 39, 30167 Hannover, Germany}
\address{$^2$Physikalisch-Technische Bundesanstalt\\
	Bundesallee 100, 38116 Braunschweig, Germany}
\address{$^3$IFN-CNR, Dipartimento di Fisica, Politecnico di Milano, Piazza L.\ da Vinci 32, Milano, 20133, Italy}
\address{$^4$Indiana University Center for Spacetime Symmetries\\
	Bloomington, IN 47405, USA}
\address{$^5$Ulmer Fundamental Symmetries Laboratory, RIKEN\\
	Hirosawa, Wako, Saitama 351-0198, Japan}

\begin{abstract}
Current precision experiments with single (anti)protons to test CPT symmetry 
progress at a rapid pace, 
but are complicated by the need to cool particles to sub-thermal energies. 
We describe a cryogenic Penning-trap setup for $^9$Be$^+$ ions 
designed to allow coupling of single (anti)protons to laser-cooled atomic ions 
for sympathetic cooling and quantum logic spectroscopy. 
We report on trapping and laser cooling of clouds and single $^9$Be$^+$ ions. 
We discuss prospects for a microfabricated trap 
to allow coupling of single (anti)protons to laser-cooled $^9$Be$^+$ ions 
for sympathetic laser cooling to sub-mK temperatures on ms time scales. 
\end{abstract}

\bodymatter

\section{Introduction}
As a result of CPT symmetry, 
particles and their antiparticles 
must have the same mass, lifetime, charge, and magnetic moment. 
The (anti)proton is an attractive candidate to test CPT symmetry 
by comparing matter--antimatter conjugates in the baryonic sector, 
complementary to tests in the lepton sector, 
e.g., 
with electrons and positrons.\cite{dehmelt_experiments_1990} 
Magnetic-moment comparisons\cite{schneider_double-trap_2017,smorra_parts-per-billion_2017} in particular 
provide a sensitive test for potential new physics.\cite{bluhm_testing_1997} 
In Penning-trap precision measurements, 
magnetic moments can be determined 
by measuring the ratio of the Larmor frequency to the free cyclotron frequency. Compared to the electron and positron, 
the bigger mass of the (anti)proton 
makes resistive cooling to the ground state of the cyclotron motion impossible. 
The resulting finite temperature complicates the measurement of the Larmor frequency 
via the continuous Stern--Gerlach effect.\cite{mooser_resolution_2013} 
Furthermore, 
systematic effects can be proportional to the oscillation amplitude of the particle
making efficient cooling to sub-thermal energies highly desirable. 
Such temperatures could be reached through sympathetic cooling 
with a laser-cooled ion. 
Implementing this in the same potential well 
could introduce additional systematic effects 
and is not possible 
for the commonly used positively charged laser-cooled ions 
together with the antiproton. 
We pursue an approach where the (anti)proton and the laser-cooled ion 
are confined in spatially separate potential wells 
and interact remotely via the Coulomb interaction.\cite{wineland_experimental_1998} This approach has already been demonstrated 
with pairs of atomic ions in radio-frequency Paul traps.\cite{brown_coupled_2011,harlander_trapped-ion_2011}

\section{Trapping single $^9$Be$^+$ ions}
We have built and commissioned a cryogenic Penning-trap system 
to implement this approach. 
The setup is based on the BASE CERN setup,\cite{smorra_base_2015} 
enhanced to allow for laser access for ablation loading, 
photo-ionization, 
Doppler and ground-state cooling of $^9$Be$^+$ ions, 
as well as for detection of laser-induced fluorescence. 
The setup is operated at a magnetic field strength of $5\,$T 
and cooled using a vibration-isolated cryocooler. 
It is described in more detail in Ref.\ \refcite{niemann_cryogenic_2019}. 
We load clouds of $^9$Be$^+$ ions 
by focusing single $532\,$nm laser pulses 
with tens of $\mu$J pulse energy 
onto a solid beryllium target embedded into a trap electrode 
via an in-vacuum off-axis parabolic mirror. 
We laser cool the ions and detect their presence 
through laser-induced fluorescence 
using a laser beam at $313\,$nm. 
We apply a series of waveforms to the trap electrodes 
in order to subsequently reduce the number of particles, 
until single ions can be observed 
as identified by discrete steps in the level of laser-induced fluorescence. 

\section{Experimental prospects for $^9$Be$^+$ ions in Penning traps}
We have recently demonstrated spin-motional control of $^9$Be$^+$ ions 
using a spectrally tailored frequency comb.\cite{paschke_versatile_2019} 
While the experimental demonstration was carried out 
at a comparatively low magnetic field in a microfabricated Paul trap, 
this type of manipulation lends itself to ground-state cooling of $^9$Be$^+$ ions 
in the $5\,$T field of the Penning trap, 
where electron-spin energy levels 
with a level spacing of $\sim\! 140\,$GHz 
need to be coupled using a stimulated Raman process. 
Our immediate plans for the setup 
comprise the demonstration of ground-state cooling 
and the coupling of pairs of $^9$Be$^+$ ions 
as a precursor for sympathetic cooling of (anti)protons. 

\section{(Anti-)protons and quantum logic spectroscopy}
\begin{figure}
\begin{centering}
	\includegraphics[width=3.5in]{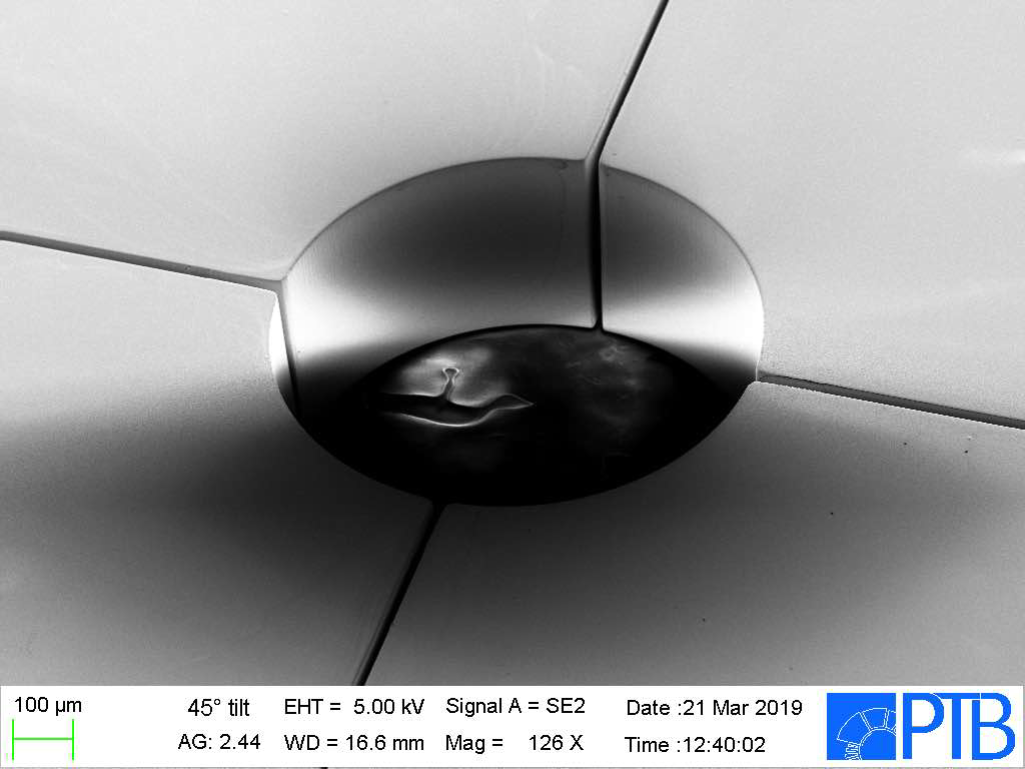}
	\caption{Gold electroplated silicon test structure for a cylindrical micro Penning trap.}
	\label{fig1}
	\end{centering}
\end{figure}

We have designed an off-axis proton source for the apparatus, 
which is awaiting commissioning. 
Coupling a single (anti)proton to a laser-cooled $^9$Be$^+$ ion 
will likely require a microfabricated trap 
in order to minimize the distance between the particles 
and obtain a strong coupling. 
Compared to the commonly used microfabricated surface-electrode Paul traps, 
a ring-shaped electrode of a micro Penning trap 
would have to be metal coated 
both on the front and back face of the disc 
and on the inside of the ring. 
We have produced test structures 
using deep reactive ion etching of a silicon wafer. 
For metalization, 
wafers have been coated 
either using resistive evaporation of Ti and Au under constant rotation of the sample
or using direct sputter deposition of Au. 
On top of these thin metal films, 
a thick electroplated film of Au has been grown and structured 
using optical lithography. 
A sample structure is shown in Fig.~\ref{fig1}. 
The process will be extended by including the spacers 
to electrically isolate multiple such rings 
already at the stage of the wafer etching. 
Electrical isolation will be possible 
through the use of the lithographic step, 
which allows us to leave parts of the sample uncoated. 
Once all of the above ingredients have been implemented 
and a suitable spin-motional coupling mechanism 
for the (anti)proton has been implemented, 
quantum logic spectroscopy\cite{heinzen_quantum-limited_1990,wineland_experimental_1998} 
could be envisioned as a means 
of probing all relevant transitions of single (anti)protons 
out of the motional ground state. 

\section*{Acknowledgments}
We acknowledge 
the support of Wissenschaftlicher Ger\"atebau at PTB 
for the construction of the setup and the rotation holder, 
the support of the PTB cleanroom team 
concerning the fabrication of the sample structures, 
and P.\ Hinze for the SEM pictures. 
We acknowledge funding from ERC StG ``QLEDS,'' 
and from DFG through SFB 1227 DQ-\textit{mat} (project B06), 
DFG grant ``Apparatur f\"ur kryogene Ionenfallen,'' 
and the clusters of excellence QUEST and Quantum Frontiers. 
RL acknowledges support from the Alexander von Humboldt Foundation.

\end{document}